\documentclass[prl,twocolumn, 10pt, preprint,nofootinbib]{revtex4}
\usepackage{graphicx}
\begin{document}

\title{Dark matter, neutron stars and strange quark matter }

\author{M. Angeles Perez-Garcia$^1$~\footnote{mperezga@usal.es}, Joseph Silk$^2$~\footnote{j.silk1@physics.ox.ac.uk} and Jirina R. Stone$^{2,3}$~\footnote{j.stone1@physics.ox.ac.uk}}

\affiliation{$^1$ Departamento de F\'{i}sica Fundamental and IUFFyM, \\Universidad de Salamanca, 
Plaza de la Merced s/n 37008 Salamanca\\
$^2$ Oxford Physics, University of Oxford, Keble Road OX1 3RH, Oxford, United Kingdom \\
$^3$ Department of Physics and Astronomy, University of Tennessee, Knoxville TN 37996, USA}

\date{\today}

\begin{abstract}

We show that self-annihilating neutralino WIMP dark matter accreted onto neutron stars may provide a mechanism to seed compact objects with long-lived lumps of strange quark matter, or {\it strangelets}, for WIMP masses above a few GeV. This effect may trigger a conversion of most of the star into a strange star. We use an energy estimate for the long-lived strangelet based on the Fermi gas model combined with the MIT bag model to set a new limit on the possible values of the WIMP mass that can be especially relevant for subdominant species of massive neutralinos.
\end{abstract}
\maketitle
There is compelling evidence \cite{rev} that most of the matter in the Universe is
dark matter (DM).
Amongst the beyond-the-Standard Model particles that may constitute DM, WIMPs (weakly interacting massive particles), typically the LSP neutralinos of SUSY models,  are the most frequently considered. 
 WIMPs from the galactic halo can be accreted onto compact massive objects \cite{press}, such as neutron stars (NS) \cite{kouvaris2008, goldman1989} and white dwarfs \cite{bertone2008, Cullough2010}.  Our focus here will be on the consequences of neutralino WIMP capture by NS. Accretion of DM has been shown \cite{kouvaris2008,delavallaz2010} to play a role in late cooling  for NS over time-scales  longer than $\tau_{\rm NS} > 10^{\rm 7}$ yr when the WIMP annihilation rate has already reached equilibrium with the accretion rate. Then the rate of released energy becomes time and temperature-independent, and  will dominate over photon  cooling processes. However this effect is  hard to detect.

The signature of WIMP self-annihilations (assuming WIMPs are Majorana particles) is energy release in the range equivalent to twice the WIMP mass, leading to subsequent particle production and/or heating of the star. Current predictions of the WIMP mass span the range from 1 $\rm GeV/c^{\rm 2}$ up to 10 $\rm TeV/c^{\rm 2}$ \cite{masswimp}. Other, more exotic, DM candidates include super-heavy WIMPzillas \cite{chung2001,kolb2007} with masses in the range $10^{\rm 10}-10^{\rm 15}$ $\rm GeV/c^{\rm 2}$ and SIMPzillas - heavy and strongly interacting \cite{faraggi2000,albuquerque2000} (see recent experimental constraints from ICECUBE \cite{icepaper}).
 
Compact objects such as black holes or NS are considered to be gravitational accretors of DM. Non-rotational NS models, yielding gravitational mass, $M$, and radius, $R$, are based on the Tolman-Oppenheimer-Volkoff (TOV) equations, given an equation of state (EoS) that relates pressure to energy density and temperature, $p=p(\epsilon, T)$. A typical NS has a mass M$ \approx$ 1.4 M$_\odot$, radius $R \approx 10\,$ km, surface temperature $T_{\rm surf} \approx 100 \,$keV and central mass density $\rho \approx 1.5 \times$ 10$^{\rm 15}$ g/cm$^{\rm 3}$. 
There are  a variety of model predictions for  NS  composition  \cite{lattimer2007} but there is as yet no consensus about the fundamental nature of the matter in NS interiors \cite{
weber2009}.

Witten \cite{Witten} proposed in 1984 that strange quark matter (SQM) made of $u$, $d$ and $s$ quarks is absolutely stable and forms the true ground state of hadronic matter, and that strange stars, made of SQM and bound by the strong interaction, should exist with very different properties from those predicted for hadronic NS \cite{itoh1970}. The mechanism of conversion of metastable, charged neutral, neutron star matter in beta equilibrium, formed due to the leptonic weak interaction $p+e \leftrightarrow n+\nu$, to a two-flavor quark matter ($ud$ matter) and subsequently to $uds$ matter is not completely understood, and a variety of models have been suggested (see e.g. \cite{alcock1986,olinto1987,olesen1994,bha2006,bombaci2007,bombaci2009}). Here we show that WIMPs may  trigger a conversion of hadronic NS matter to SQM through an external mechanism of seeding the NS with long-lived lumps of SQM  as the result of energy release from DM self-annihilations in a distribution of gravitationally-captured WIMPs in the NS core.
 
The accretion rate of WIMPs captured by a typical NS is given by 
\cite{press, gould, kouvaris2008}
\begin{equation}
{\cal F}= \frac{3.042 \times 10^{25}}{ m_{\chi} (\rm GeV)} \frac{\rho_{DM}}{\rho_{DM,0}}\, (s^{-1}),
\label{rate}
\end{equation}
assuming a WIMP-nucleon interaction cross section $\sigma_{\chi n} > 10^{-45}$ cm$^{\rm 2}$. We assume that the local DM mass density in the vicinity of the NS in an average galaxy  is similar to that in the solar neighbourhood,  about $\rho_{\rm DM,0}=0.3\, \rm GeV/cm^{\rm 3}$, and that all incoming WIMPs undergo one or more scatterings while inside the star. Here m$_\chi$ is the mass of the WIMP. Due to competing effects of annihilation and evaporation, the number of accreted WIMPS at time $t$ is obtained by solving the differential equation
\begin{equation}
{\dot N}= {\cal F}-\Gamma_{\rm annih}-\Gamma_{\rm evap}, 
\end{equation}
where $\Gamma_{\rm annih}$ is the self-annihilation rate calculated as $\Gamma_{\rm annih}=\langle\sigma_{\rm annih} v\rangle \int n^2_{\chi} dV=C_{\rm A}N(t)^{\rm 2}$, $\langle\sigma_{\rm annih}v\rangle$ is the product of thermally-averaged WIMP self-annihilation cross section and velocity, and  $n_\chi=\frac{\rho_{\rm DM}}{m_{\chi}}$, the number density of WIMPs inside the NS, is assumed to be constant. The evaporation rate, $\Gamma_{\rm evap}$, decays exponentially with temperature  $\sim e^{\rm -G M m_{\chi}/R T}$,  and is negligible with respect to the annihilation rate for NS  with internal T $\sim$ 0.1 MeV \cite{krauss1986}. With this simplification, the population of WIMPs at time $t$ is given by
\begin{equation}
N(t)= ({\cal F} \tau) {\rm tanh}(t/ \tau),
\end{equation}
where the time-scale is $\tau=1/{\rm \sqrt{{\cal F }C_{\rm A}}}$. For $t>>\tau$, when the equilibrium between accretion and annihilation has been reached, the number of particles accreted is time-independent, $N={\mathcal F}\tau$.

Assuming a regime when velocities and positions of the WIMPs follow a Maxwell-Boltzmann distribution with respect to the centre of the NS, the thermalisation volume in the compact star has a radius \cite{bertone2008} $r_{\rm th}= (\frac{9 k T_c}{4 \pi G \rho_c m_{\chi}})^{1/2}=64 \, (\frac{T}{10^{\rm 5} K})^{1/2} (\frac{10^{\rm 14} {\rm g/cm^3}}{\rho_{\rm c}})^{1/2} (\frac{100 \,{\rm GeV/c^{\rm 2}}}{m_{\chi}} )^{1/2}\,(\rm cm)$. 
Taking typical NS conditions as central internal temperature $T_c=10^{\rm 9}$ K, $\rho_{\rm c}$=10$^{\rm 15}$ \rm g/cm$^{\rm 3}$ and m$_{\chi}=1\, \rm GeV/c^2-10\, TeV/c^2$, r$_{\rm th}\approx 2\times 10^{5}-2 \times 10^{2}\,{\rm cm}$. Then the thermalization volume is  $V_{\rm th}=\frac{4}{3} \pi r_{\rm th}^{\rm 3} \approx  (10^{\rm -2}-10^{\rm -8})\, V_{\rm NS}$.

The exact energy released due to WIMP self-annihilation is dependent on the nature of WIMPs and on the output  product channel and has to be calculated separately for each case. As we are interested in order of magnitude estimates, we  assume that after an annihilation occurs, a fraction of this energy, given by the efficiency factor \cite{ffactor}, $f$, is deposited locally. The rate of energy released in annihilation processes in the star in the thermal equilibrium volume can be expressed as
\begin{equation}
{\dot E_{\rm annih}} = C_A N^2 m_{\chi}c^2 = f {\cal F} m_{\chi}c^2,
\label{edot}
\end{equation}
where $f \approx 0.01-1$. If most of the total energy is not converted into neutrinos, then $f$ is close to unity. 

Eq.(\ref{edot}) includes the option that  $\rho_{DM}$, in the vicinity of the NS, may differ from the local standard  DM density, $\rho_{\rm DM,0}$. 
Taking $\rho_{DM}$=$\rho_{\rm DM,0}$ and an efficiency rate $f=0.9$, we have that in the range $m_{\chi} \approx 1-10^4$ $\rm GeV/c^{\rm 2}$
\begin{equation}
{\dot E_{\rm annih}}=2.74 \times 10^{\rm 25}- 10^{\rm 29}\,{\rm GeV/s}.
\end{equation}
The energy released will convert partly into heat, which can stimulate $u,d$ quark bubble formation by thermal fluctuations in metastable hadronic matter via strong interactions \cite{olesen1994}. The timing and conditions of these transitions depend strongly on the pressure in the central regions of the star given by the EoS of hadronic matter. Then $ud$ matter undergoes non-leptonic weak reactions, such as $u+d \leftrightarrow u+s$, to form drops of strange $uds$ matter, which has lower energy as a result of the reduction in Fermi energies through the introduction of a new flavor. The stability, among other properties of these drops, commonly called {\it strangelets}, has been extensively studied (see e.g. \cite{sqm}) and depends on the electrical charge, strangeness fraction and size. The energy needed to form a stable 
long-lived strangelet of baryonic number $A$ can be calculated from its quark constituents, in a first approximation, using either the  MIT bag model with shell mode filling or the liquid drop model (for details see e.g. \cite{madsen2,schaffner}). The former approach is important particularly at low $A$  but becomes impractical for large $A$. Minimum values of $A$ and lifetimes of strangelets have been calculated in the MIT bag model approach \cite{schaffner} at zero temperature for different values of the bag constant, $B$, charge and strangeness. Finite temperature effects on strangelets  have also been examined \cite{zhang}. Although they do not appreciably modify the minimum baryon number of a strangelet, they shift stability conditions towards the high negative electrical charge fraction and high strangeness.

In the ideal Fermi-gas approximation, the binding energy of a strangelet with baryonic number $A$ composed of massive quarks of flavor $i$ $(i=u,d,s)$ is \cite{madsen2} 
\begin{equation}
E^A(\mu_i, m_i, B) =  \sum_i (\Omega_i + N_i \mu_i)+BV 
\end{equation}
where $N_i $ is the quark number in the strangelet of baryon number $A = \frac{1}{3} (N_{\rm u}+N_{\rm d}+N_{\rm s})$, $\mu_i(n_{\rm A})$ is the quark chemical potential at baryonic number density $n_{\rm A}=\frac{1}{3} \displaystyle \sum_i n_i$. $n_i$ are the $ith$-quark number densities in the strangelet. The thermodynamical potential for the $ith$-quark type with mass  $m_i$ is given by $\Omega_i (\mu_i, m_i)=\Omega_{i ,V}V+ \Omega_{i, S} S +\Omega_{i ,C} C$ being $V=A/n_A=\frac{4}{3} \pi R^3$ the volume, $S=4\pi R^2$ the surface and $C=8 \pi R$ the curvature of the spherical strangelet. Expressions for these potentials can be obtained from \cite{madsen2}. The masses of the quarks are taken as $m_u=2.55$ MeV, $m_d=5.04$ MeV, $m_s=104$ MeV respectively \cite{masswimp}.  

In order to consider charged strangelets a Coulomb correction term can be added as $E_{\rm coul}=\frac{4}{3}(\frac{\alpha Z^2_V}{10R} +\frac{\alpha Z^2}{2R})$ with $Z_V=\displaystyle \sum_i q_i n_i V$ and $Z=\displaystyle \sum_i q_i$ given the $ith$-quark charge $q_i$. 
\begin{figure}[hbtp]
\begin{center}
\includegraphics [angle=0,scale=1.0] {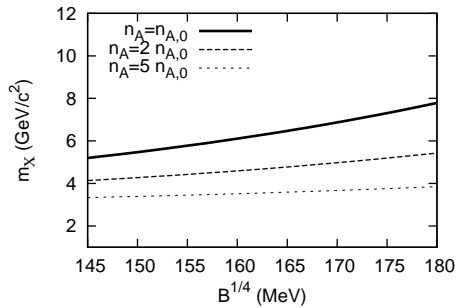}
\caption{ Mass of the DM WIMP as a function of $B^{1/4}$ for different values of the baryonic density, $n_A,$ taking the  efficiency factor  $f=0.9$.}
\label{Fig1}
\end{center}
\end{figure}
Then, the energy of a strangelet with baryon number $A$ is
\begin{equation}
E^A_{\rm slet}\approx E^A(\mu_i, m_i, B)+E_{\rm coul}.
\end{equation}
Note that we have neglected a zero-point motion correction as done in \cite{madsen2}. Estimation of the minimum value of $A$ in a long-lived strangelet is model-dependent but typical minimum $A$ values are in the range $A_{\rm min} \approx 10-600$ \cite{alford2006}. Schaffner et al. \cite{schaffner} find strangelets with `magic' numbers of quarks for bag constant values $145 \leq B^{1/4}  \leq 170$ MeV. If smaller strangelets than the minimum $A$ are created, they will decay rapidly. However, it remains for further investigation to see whether quark drop percolation plays a significant role here. This may happen as a result of multiple WIMP annihilations over scales of several fm  (typical size of a strangelet). A long-lived cluster may be formed from different small clusters with interdistance $l$ on the diffusion time-scale, $t_{\rm diff}\sim l/v_{\rm diff}$, before they decay over a mean lifetime, $\tau \sim 10^{\rm -10}-10^{\rm -5}$ s. This effect would  enhance the seeding scenario presented here by increasing the energy available for strangelet formation. The more elaborate calculations  \cite{schaffner} show that among all possible metastable strangelets, the long-lived ones have lifetimes of days.
Since this time-scale is larger than the conversion time-scale $t_{\rm conv} \approx 100\,s $ \cite{bha2006}, it is in principle possible that strange stars could be formed if this conversion is triggered. As an estimate, the rate of formation of long-lived strangelets with, for example, $A=10$ would be
\begin{equation}
{\dot N_{\rm slet}}={\dot E_{annih}}/E^A_{\rm slet}. \end{equation}
For $f=0.9$, $n_A=n_{A,0}$, $m_{\chi}=1$ $\rm GeV/c^{\rm 2}$, a number ${\dot N_{\rm slet}} \approx 10^{\rm 23}  \,s^{-1}$ are created. Assuming that in the center of the star,  self-annihilations are dominant, the energy deposited in the medium, per annihilation,  will be $2 f m_{\chi}c^2$. This allows us  to set a limit on the energy scale, at a given central baryonic density $n_A$. 
\begin{equation}
2 f m_{\chi}c^2 \geq E^A_{\rm slet}(\mu_i (n_A), m_i, B).
\label{ineq}
\end{equation}
Note that the mass of the WIMP can be related to the strong interaction through the  MIT bag constant. Eq.(\ref{ineq}) provides a strong limit to complement current estimates  from direct  \cite{direct} as well as indirect searches. 
In Fig.\ref{Fig1} we show the WIMP mass as function of $B^{1/4}$ for a long-lived strangelet baryon number $A=10$ and an efficiency rate $f=0.9$ for $n_A=n_{A,0}$ (solid line), $n_A=2 n_{A,0}$ (long dashed line) and $n_A=5 n_{A,0}$ (short dashed line), with $n_{A,0}=0.17 fm^{-3}$. It can be seen that there is a weak dependence on nuclear physics input, parameterized by $B$. As the central baryon density increases, the lower limit on the WIMP mass decreases below  $\sim \rm 4\, GeV/c^2$. We adopt  $n_A=5 n_{A,0}$ as the highest value of the central baryon density.

\begin{figure}
\begin{minipage}[b]{1.0 \linewidth} 
\centering
\includegraphics[scale=1.0]{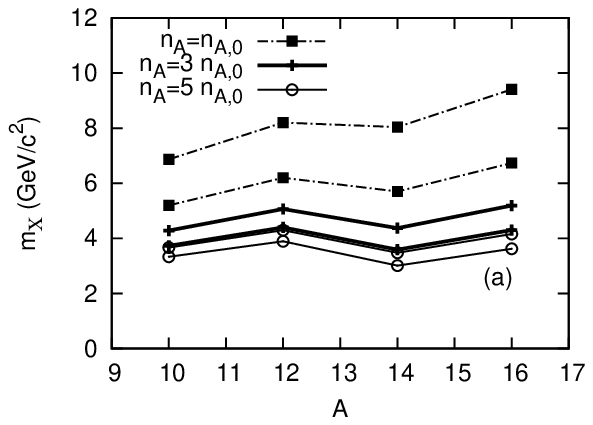}
\end{minipage}
\hspace{0.1cm} 
\begin{minipage}[b]{1.0 \linewidth}
\centering
\includegraphics[scale=1.0]{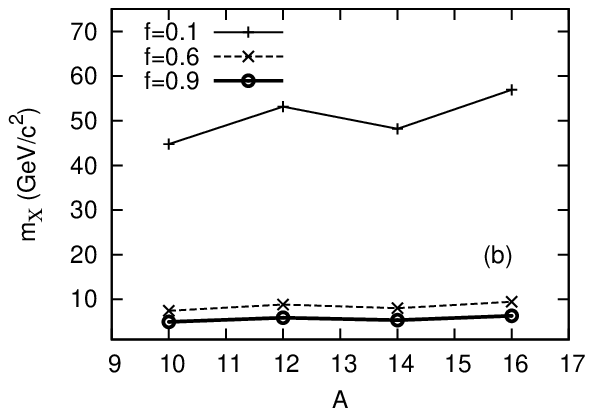}
\caption{WIMP mass as a function of long-lived strangelet baryon number $A$ for (a) limiting values of $B^{1/4}=140, 170$ MeV, and $n_A/n_{A,0}=1,3,5$ (b) $B^{1/4}=170$ MeV, $3n_{A,0}$ and $f=0.1, 0.6, 0.9$.}
\label{Fig23}
\end{minipage}
\end{figure}

In Fig. \ref{Fig23}, we show (a) the WIMP mass as a function of $A$ in the long-lived strangelet for an efficiency rate $f=0.9$ and different values of the central density $n_A=n_{A,0}$
(dash-dotted line), $3 n_{A,0}$ (thick solid line), $5 n_{A,0}$  (thin solid line)  for limiting values of the bag constant. For each  pair of curves, the lower line is for $B^{1/4}=145$ MeV and the  upper line is for $B^{1/4}=170$ MeV. For a typical baryonic density of $\sim 3 n_{A,0}$, a lower linit for the WIMP mass $m_{\chi}\gtrsim 4$ $\rm GeV/c^{\rm 2}$ is predicted.
In Fig.\ref{Fig23} we show in (b) the WIMP mass as a function of the long-lived strangelet baryon number $A$ in the range $10-16$ according to ref.\cite{schaffner} for different values of the efficiency rate $f$ and $B^{1/4}=145$ MeV and $n_A=3 n_{A,0}$. 
We consider $f=0.1$ (thin solid line), $f=0.6$ (dashed line) and $f=0.9$ (thick solid line). 
 As more massive WIMPs annihilate, the smaller the efficiency rate  becomes that is required to trigger a conversion from nucleon to strange quark matter. 

In summary, through the mechanism of self-annihilation of DM candidates in the central regions of a typical NS proposed in this work, we have derived a limit on the mass of DM particle candidate using the fact that there is a minimum long-lived strangelet mass needed to trigger a conversion from nuclear matter into strange quark matter. In the range $m_{\chi}\gtrsim 4$ $\rm GeV/c^2$ the energy released in WIMP self-annihilation is sufficient to burn nucleon matter into a long-lived strangelet that will trigger full conversion to a strange star. It is important to emphasize that so far the lack of knowledge of DM properties prevents us from making any definite statement on the WIMP mass other than excluded regions, but this work presents a scenario compatible with current experimental direct and indirect searches. Two important pieces of observational information would provide supporting evidence for the mechanism of NS conversion proposed here and potentially improve the derived limit on the WIMP candidate: (i) observation of a quark star and measurement of its mass and radius, which could yield constraints on its EoS; (ii) identification of formation of strangelets and its properties at LHC or RHIC.

In the astrophysical scenario, one consequence of the NS conversion process could be related to the emission of $\gamma$-ray bursts (GRB). According to Ma et al. \cite{ma} the energy release in a Super-Giant-Glitch due to the change in radius of an initial to final star configuration is $\Delta E \sim 10^{53} \frac{\Delta R}{R}$ erg. The calculated rate of GRBs in the galaxy is $R_{\rm GRB} \sim 10^{-6}\, {\rm yr}^{-1}$ in good agreement with observations. Such a huge burst would happen only once in the lifetime of the NS. 
Bombaci et al. \cite{bombaci2006}, pointed out that the energy liberated during the NS conversion process will be carried out mainly by neutrinos that will be visible near the surface as GRBs. In addition, the strong magnetic field in the NS can cause collimated GRBs and the rotation of the NS could be a source of anisotropic emission. They have also investigated the effect of uncertainties introduced by different EoS. Recent measurements of the source of short GRB 090510 by Fermi LAT report an energy  $E=(1.08 \pm 0.06 ) \times 10^{53} \, \rm erg$ in the (10 keV-30 GeV) range \cite{fermilat} and a time duration of most of the signal, $T_{\rm 90}=0.6-9$ s, compatible with the expected energy release \cite{bombaci2006} and duration of the conversion of a hadronic to quark star \cite{bha2006}. 

An interesting open question is whether conventional pulsar glitches observed so far only in hadronic stars would be possible in quark stars which are predicted to have no crust or very thin one. Some models \cite{alford2006} suggest that it is unlikely that quark stars would give rise to glitches as these require a crust with charged strangelets inmersed both in electron gas and a superfluid and there is no superfluid in the crust of quark stars. However if the quark matter in the core of the star were to exist in crystalline color-superconducting phase, glitches might origite there.

The rate of conversions of hadronic to strange NS via the mechanism presented in this work is directly related to the DM density, since it depends on the galactic DM profile. Thus this rate could be potentially observable as a function of position of the NS in the galaxy. If WIMP annihilation is considered as an effective way of seeding NS with strangelets, multicomponent DM may have observable consequences  in the conversion of NS to strange stars even if the annihilating component (e.g. SIMPzillas) is subdominant, as shown by the ICECUBE constraint.
 It would be also interesting to explore whether the annihilation of DM could cause changes in superfluid phases of matter inside NS and its effect on cooling and rotation patterns. 

We would like to thank J. Bell and J. Miller for helpful discussions. M.A.P.G. would like to thank University of Oxford for kind hospitality. She is a member of the COMPSTAR collaboration and acknowledges partial support under MICINN project FIS-2009-07238  and Junta de Castilla y Le\'on GR-234. 


\begin{thebibliography}{9}
%
\bibitem{rev}
G. Bertone, D. Hooper, and J. Silk, Phys. Rep. 405, 279 (2005).
%
\bibitem{press}
W. H. Press and D. Spergel, ApJ 296 679 (1985)
%
\bibitem{kouvaris2008}
C. Kouvaris, Phys.Rev. D 77, 023006 (2008)
\bibitem{goldman1989}
I. Goldman and S. Nussinov, Phys. Rev. D 40 3221 (1989)
%
\bibitem{bertone2008}
G. Bertone and M. Fairbairn, Phys. Rev. D 77, 043515 (2008)
\bibitem{Cullough2010}
M. McCullough and M. Fairbairn, Phys. Rev. D 81, 083520 (2010) 
%
\bibitem{delavallaz2010}
A. de Lavallaz and M. Fairbairn, arXiv: 1004.0629v1 (2010), C. Kouvaris and P. Tinyakov arXiv:1004.0856 [astro-ph.GA]
%
\bibitem{masswimp}
S. Eidelman et al. (Particle Data Group), Phys. Lett. B 592, 1 (2004)
\bibitem{chung2001}
D. J. H. Chung et al., Phys. Rev. D 64, 043503 (2001)
%
\bibitem{kolb2007}
E. W. Kolb, A. A. Starobinsky, I. I. Tkachev, JCAP07,005 (2007)  
%
\bibitem{faraggi2000}
 A. E. Faraggi and M. Pospelov, Astropart. Phys. 16, 451
(2002) [arXiv:hep-ph/0008223]
%
\bibitem{albuquerque2000}
I. F. M. Albuquerque, L. Hui and E.W. Kolb, Phys. Rev.
D 64 (2001) 083504 [arXiv:hep-ph/0009017].
%
\bibitem{icepaper}
I. Albuquerque and C. Perez de los Heros [arXiv:astro-ph/1001.1381]
%
%
\bibitem{lattimer2007}
J.M. Lattimer and M. Prakash, Phys. Rep. 422, 109 (2007)
\bibitem{weber2009}
F. Weber, R. Negreiros and P. Rosenfield, in {\it Neutron Stars and Pulsars}, Springer Berlin Heidelberg, p. 213, 2009.
\bibitem{Witten}
E. Witten, Phys. Rev. D 30, 272 (1984)
%
\bibitem{itoh1970}
N. Itoh, Prog. Theor. Phys., 44 291 (1970), A. Kurkela et al Phys. Rev. D 81 105021 (2010)  
%
\bibitem{alcock1986}  
C. Alcock, E.Farhi and A. Olinto, ApJ 310, 26 (1986)
\bibitem{olinto1987}
A.V.Olinto, Physics Letters B192, 71 (1987)
\bibitem{olesen1994}
M.L.Olesen and J.Madsen, Phys. Rev. D49, 2698 (1994)
%
\bibitem{bha2006}
A. Bhattacharyya et al.,Phys. Rev. C 74, 065804 (2006)
\bibitem{bombaci2007}
I. Bombaci, G. Lugones, and I. Vidana, Astron. Astrophys. 462, 1017 (2007)
\bibitem{bombaci2009}
I. Bombaci et al., Phys. Lett. B680, 448 (2009)
%
\bibitem{gould}
A. Gould, ApJ 321, 571 (1987)7
%
\bibitem{krauss1986}
L. M. Krauss, M. Srednicki and F. Wilczek, Phys. Rev. D 33, 2079 (1986).
%
\bibitem{ffactor}
T. R. Slatyer, N. Padmanabhan and D. Finkbeiner, Phys. Rev. D 80, 043526 (2009)
%
\bibitem{sqm}
E. P. Gilson and R. Jaffe, PRL 71, 332 (1993); J. Madsen, Phys. Rev. Lett. 85, 4687 (2000)
%
\bibitem{madsen2}
J Madsen, Phys. Rev. D 47, 5156 (1993); Phys. Rev. D 50, 3328 (1994)
%
\bibitem{schaffner}
J. Schaffner-Bielich et al., Phys. Rev. C55 3038 (1997)
%
\bibitem{zhang}
 Y. Zhang and R.-K. Su, Phys. Rev. C67, 015202 (2003)
%
\bibitem{alford2006}
M. G. Alford et al., Phys.Rev. D73, 114016 (2006)
\bibitem{direct}
J. Kopp, T. Schwetz and J. Zupan, [arXiv:astro-ph/0912.4264]
%
\bibitem{ma}
F. Ma and B. Xie, ApJ 462, L63 (1996)
%
\bibitem{bombaci2006}
I. Bombaci and B. Datta, ApJ 530, L69 (2000)
%
\bibitem{fermilat}
M. Ackermann et al., ApJ 716, 1178 (2010)
%
\end{thebibliography}
\end{document}